\documentclass[a4paper,11pt]{article}
\usepackage{pos}

\title{Illuminating early-stage dynamics of heavy-ion collisions through photons at RHIC BES energies}
\ShortTitle{Photons at the RHIC BES}

\author*[a, b]{Chun Shen}
\author[c]{Abel Noble}
\author[d,e]{Jean-Fran\c{c}ois Paquet}
\author[f]{Bj\"{o}rn Schenke}
\author[g]{Charles Gale}

\affiliation[a]{Department of Physics and Astronomy, Wayne State University, Detroit, Michigan 48201, USA}

\affiliation[b]{RIKEN BNL Research Center, Brookhaven National Laboratory, Upton, NY 11973, USA}

\affiliation[c]{Department of Physics, University of Michigan, Ann Arbor, Michigan 48109, United States}

\affiliation[d]{Department of Physics and Astronomy, Vanderbilt University, Nashville TN 37235}

\affiliation[e]{Department of Mathematics, Vanderbilt University, Nashville TN 37235}

\affiliation[f]{Physics Department, Brookhaven National Laboratory, Upton, NY 11973, USA}

\affiliation[g]{Department of Physics, McGill University, 3600 University Street, Montreal, QC, Canada H3A 2T8}

\emailAdd{chunshen@wayne.edu}

\abstract{
Heavy-ion collisions at $\sqrt{s_\mathrm{NN}} \sim 10$\,GeV probe the QCD phase diagram at large baryon densities. Because the longitudinal Lorentz contraction is small at these collision energies, understanding the dynamics during the early phase of the collision is essential for the subsequent modeling of the system evolution and for constraining the QGP transport properties at finite baryon densities. Direct photons provide undistorted information on early-stage dynamics. We model relativistic heavy-ion collisions at RHIC Beam Energy Scan energies with a hybrid dynamical approach consisting of a 3D-Glauber initial state followed by viscous hydrodynamics and hadronic transport (MUSIC + UrQMD). The implemented thermal photon emission takes into account the enhancement from finite baryon chemical potentials. We show that direct photon spectra and their anisotropic flow coefficients have a strong sensitivity to the early stage of heavy-ion collisions. Thus, they provide constraints on QGP dynamics complementary to those obtained from hadronic observables.
}

\FullConference{% HardProbes2023\\
 26-31 March 2023 \\
 Aschaffenburg, Germany\\}

\begin{document}
\maketitle

\section{Introduction}

Relativistic heavy-ion collisions open up a fascinating realm of strong many-body interactions that offers unique insights into the fundamental properties of matter under extreme conditions~\cite{Arslandok:2023utm}. These high-energy collisions create a deconfined state of matter known as the quark-gluon plasma (QGP). The QGP behaves like a fluid with minimal viscosity. It is believed to be the universe's primordial state, existing briefly after the Big Bang before transitioning into the particles we observe today. Understanding the dynamics of the QGP formation and its subsequent evolution is one of the primary goals of relativistic heavy-ion physics. While a plethora of observables has been employed to probe the properties of the QGP, such as hadronic particles and jets, recent advancements have shown that photons emitted during the early stages of the collision provide a unique and powerful tool for studying the initial conditions and subsequent evolution of the QGP~\cite{Gale:2021emg, Geurts:2022xmk}.

Photons %, as electromagnetic radiation, 
are produced at various stages of the heavy-ion collision process~\cite{Shen:2016odt, David:2019wpt}. Those generated in the initial stages carry crucial information about the early dynamics, as they are  almost unaffected by the later stages of the collision. These early photons, originating from the interaction of quarks and gluons in the QGP, offer independent access to the initial temperature, hydrodynamic flow, and other essential properties characterizing the early stages of the collision, complementary to hadronic observables~\cite{Gale:2021emg}. Through future precise measurements of photon spectra, azimuthal anisotropies, and correlations with other particles, valuable phenomenological constraints can be deduced on the temperature, thermalization time, and collective behavior of the QGP, shedding light on the crucial moments following immediately after the collision.

Moreover, the advent of novel experimental techniques and theoretical advancements has expanded the scope of photon measurements, enabling the study of more elusive phenomena within the QGP. For instance, efforts are underway to study chemical equilibration and the QGP dynamics at finite baryon density using photons as sensitive probes~\cite{Gale:2021emg, Gale:2018vuh}. Understanding the behavior of matter at finite baryon density is of utmost importance for comprehending the phase structure of quantum chromodynamics (QCD) and the transition between different phases of matter. Additionally, studies involving both real photons and dileptons can provide a comprehensive picture of the evolving system and further deepen our understanding of the QGP's intricate properties~\cite{Rapp:2016xzw, Coquet:2021gms}. 

This proceeding will highlight our recent effort of using both photon and hadronic observables to study the dynamics of relativistic heavy-ion collisions at the RHIC Beam Energy Scan (BES) energies.

\section{The Hybrid Theoretical Framework}
\label{Sec:Model}

In this work, we employ the iEBE-MUSIC hybrid framework to simulate the full (3+1)D dynamical evolution of relativistic heavy-ion collisions at the RHIC BES energies~\cite{Shen:2022oyg, Zhao:2022ayk}. As we switch among different numerical models to describe the system's dynamics in different phases, we ensure the system's energy-momentum tensor is continuously matched at the switching boundaries. The hybrid simulations provide a continuous time evolution of the local energy-momentum tensor, which is folded with the thermal photon emission rate to compute thermal photon productions from these collision events. 
To make apples-to-apples comparisons with the experimental measurements, we define the collision centrality with the final-state charged hadron multiplicity from the same rapidity region as the experiments. The anisotropic flow coefficients of thermal photons are computed with reference flow vectors determined by the charged hadrons.

\subsection{Dynamical initialization of heavy-ion collisions at intermediate collision energies}

Because the Lorentz contraction of the incoming nuclei becomes weak as the collision energy goes down, we need to simulate the dynamics of the collision system during the two incoming nuclei pass through each other. For example, the overlapping time between the incoming nuclei can reach up to about 2 fm/$c$ for central Au+Au collisions at $\sqrt{s_\mathrm{NN}}=19.6$\,GeV. This early phase can be modeled by the dynamical initialization scheme~\cite{Shen:2017ruz},
\begin{equation}
    \partial_\mu T^{\mu \nu} = 0\quad \mbox{ and }\quad \partial_\mu J^\mu_B = \rho_B.
    \label{Eq:hydroEOM}
\end{equation}
Eq.~\eqref{Eq:hydroEOM} ensures the system's energy, momentum, and net baryon charges are continuously mapped from the initial-state model to the macroscopic hydrodynamic fields.
In this work, we use the 3D MC-Glauber model~\cite{Shen:2017bsr, Shen:2022oyg} to provide the space-time distribution of energy-momentum currents $J^\mu$ and net baryon charge $\rho_B$ sources. This initial-state model dynamically simulates the energy-momentum loss of the participating nucleons or quarks as the two nuclei pass through each other. See Ref.~\cite{Shen:2017bsr, Shen:2022oyg} for details.

\begin{figure}[ht!]
    \centering
    \begin{tabular}{cc}
    \includegraphics[width=0.47\linewidth]{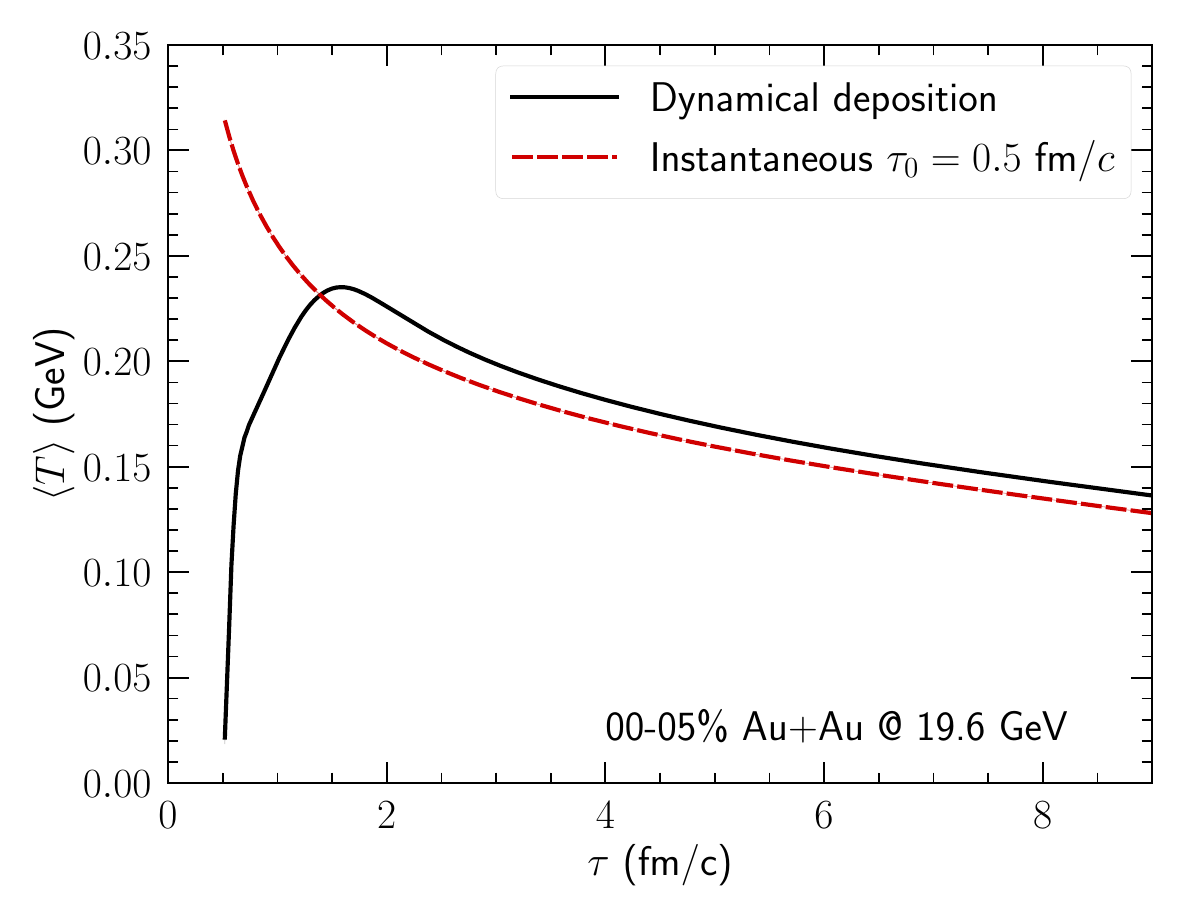} & 
    \includegraphics[width=0.47\linewidth]{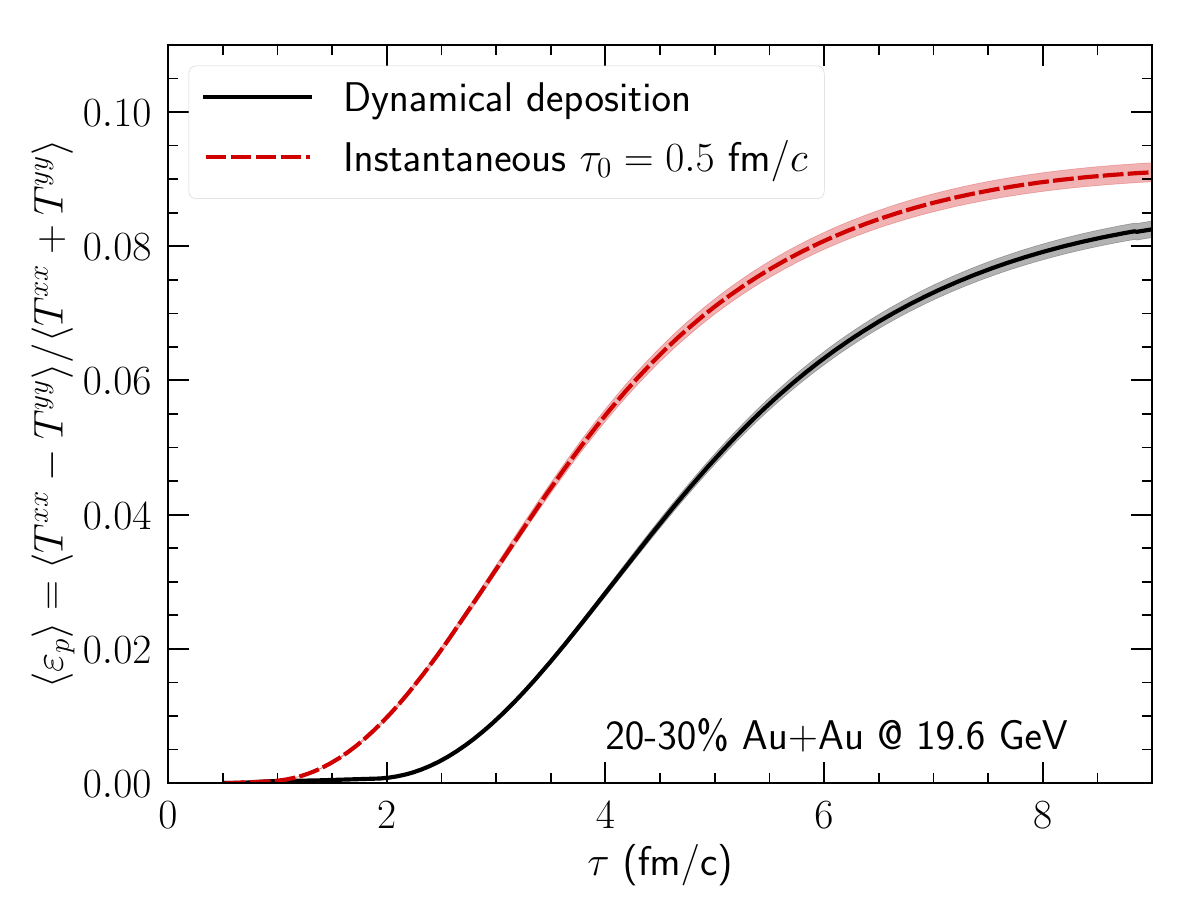} \\
    \includegraphics[width=0.47\linewidth]{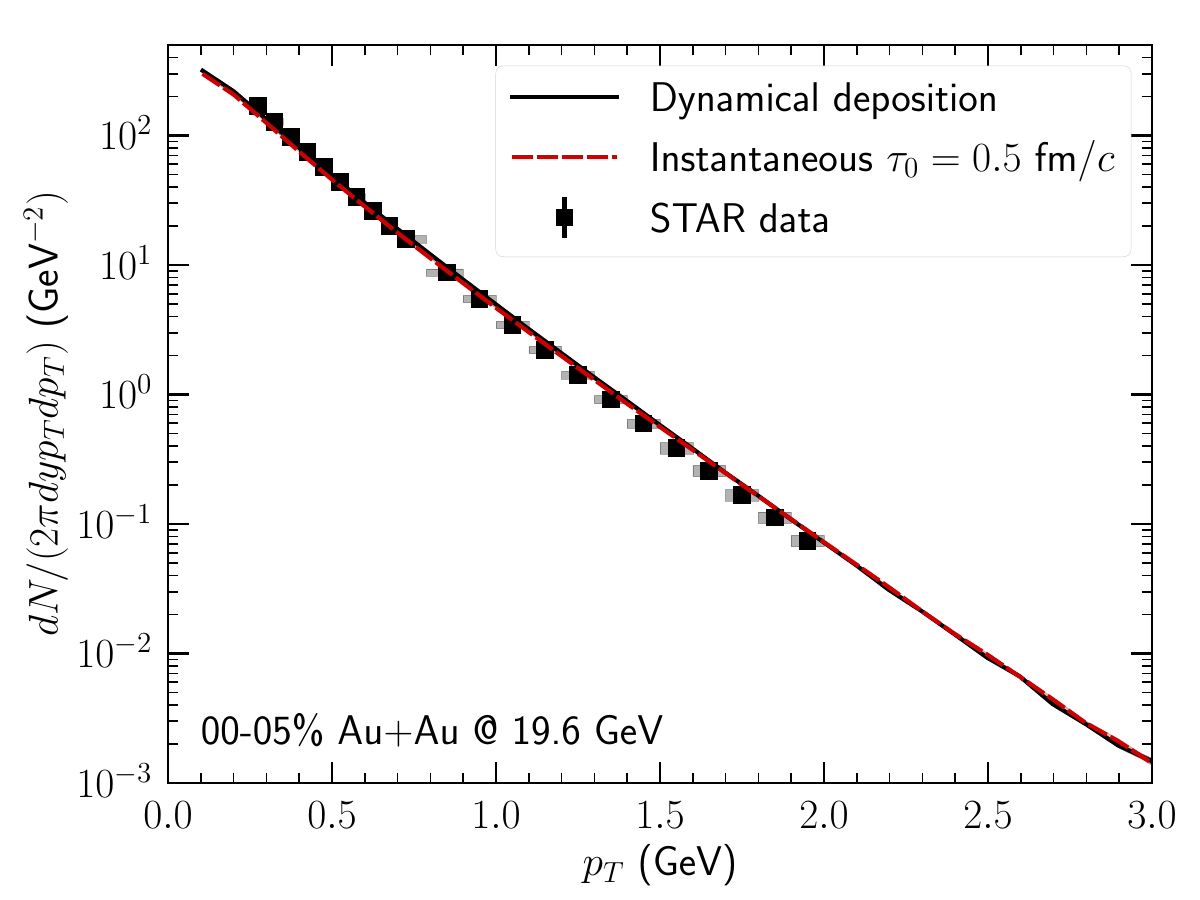} & 
    \includegraphics[width=0.47\linewidth]{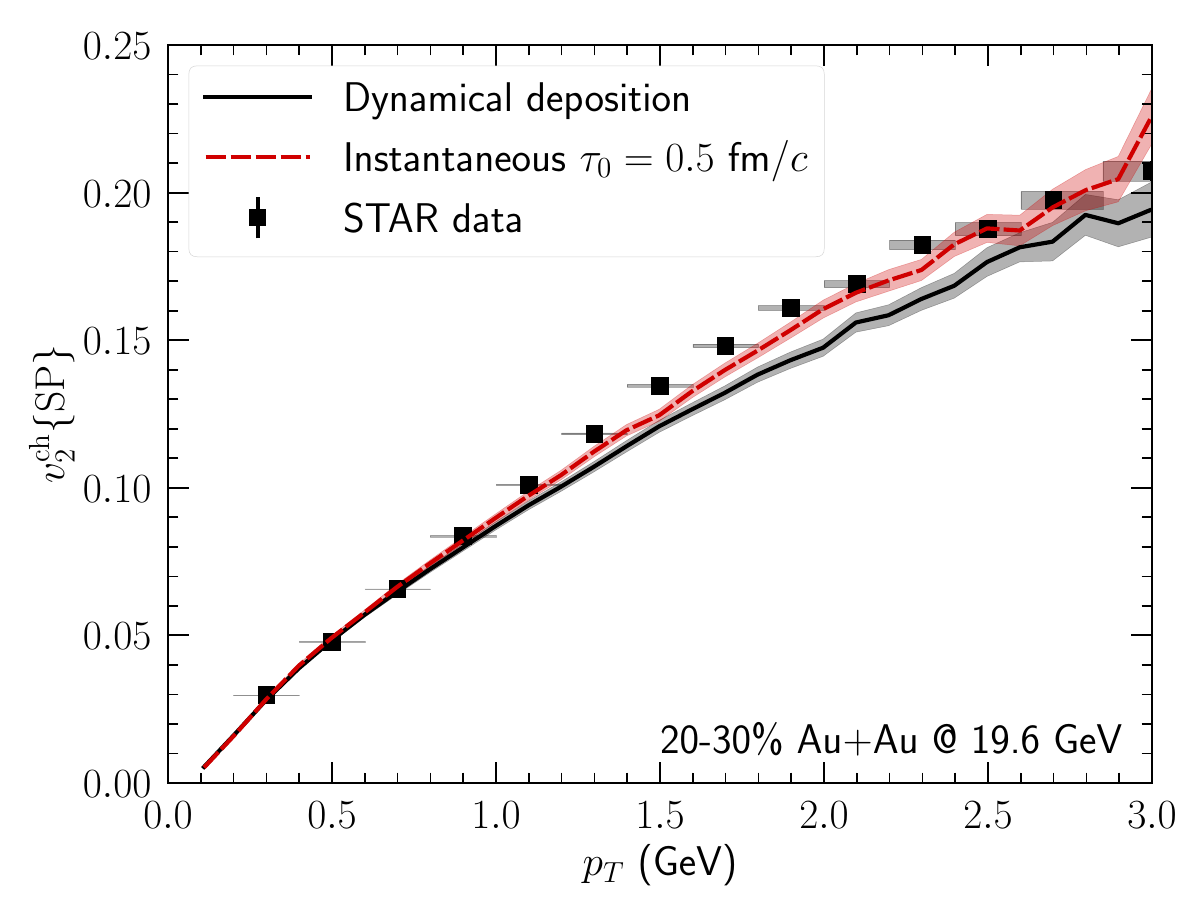} \\
    \includegraphics[width=0.47\linewidth]{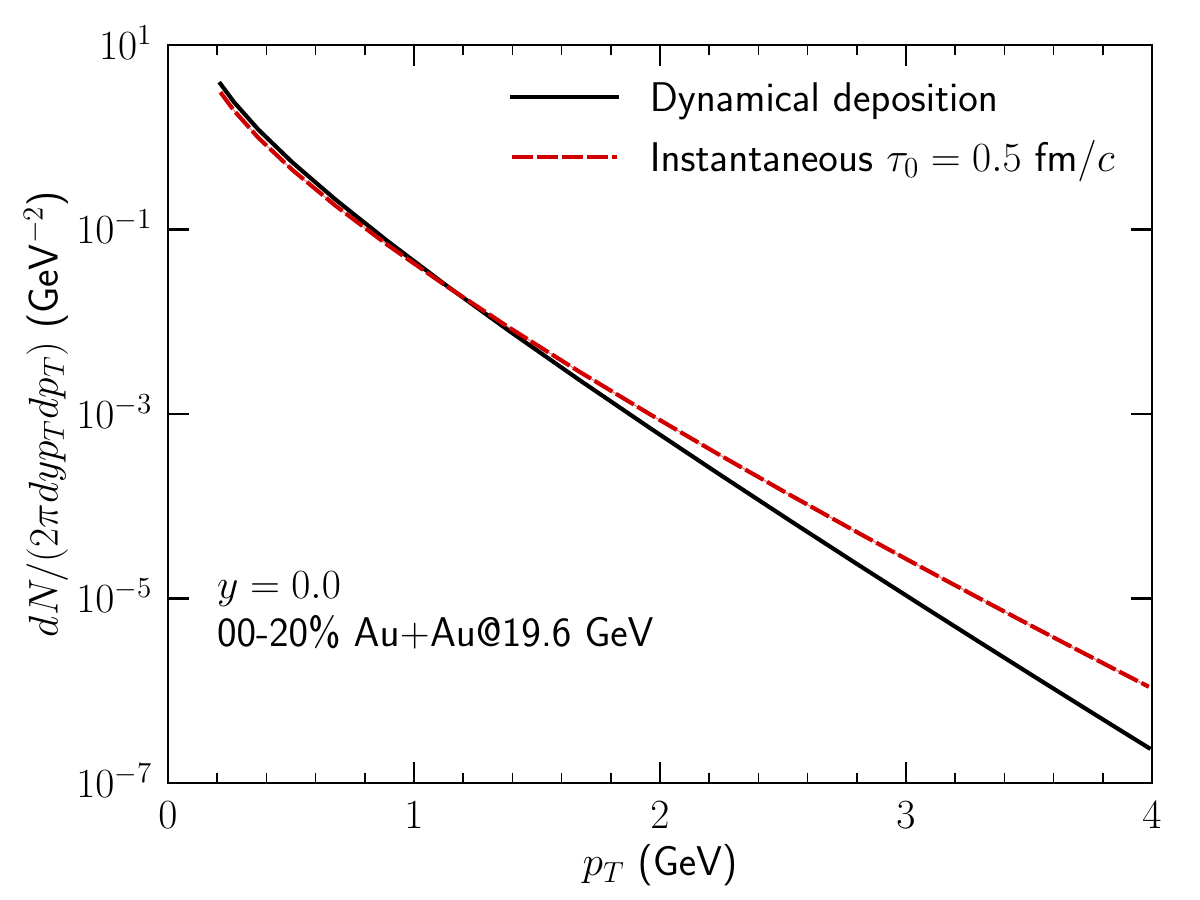} & 
    \includegraphics[width=0.47\linewidth]{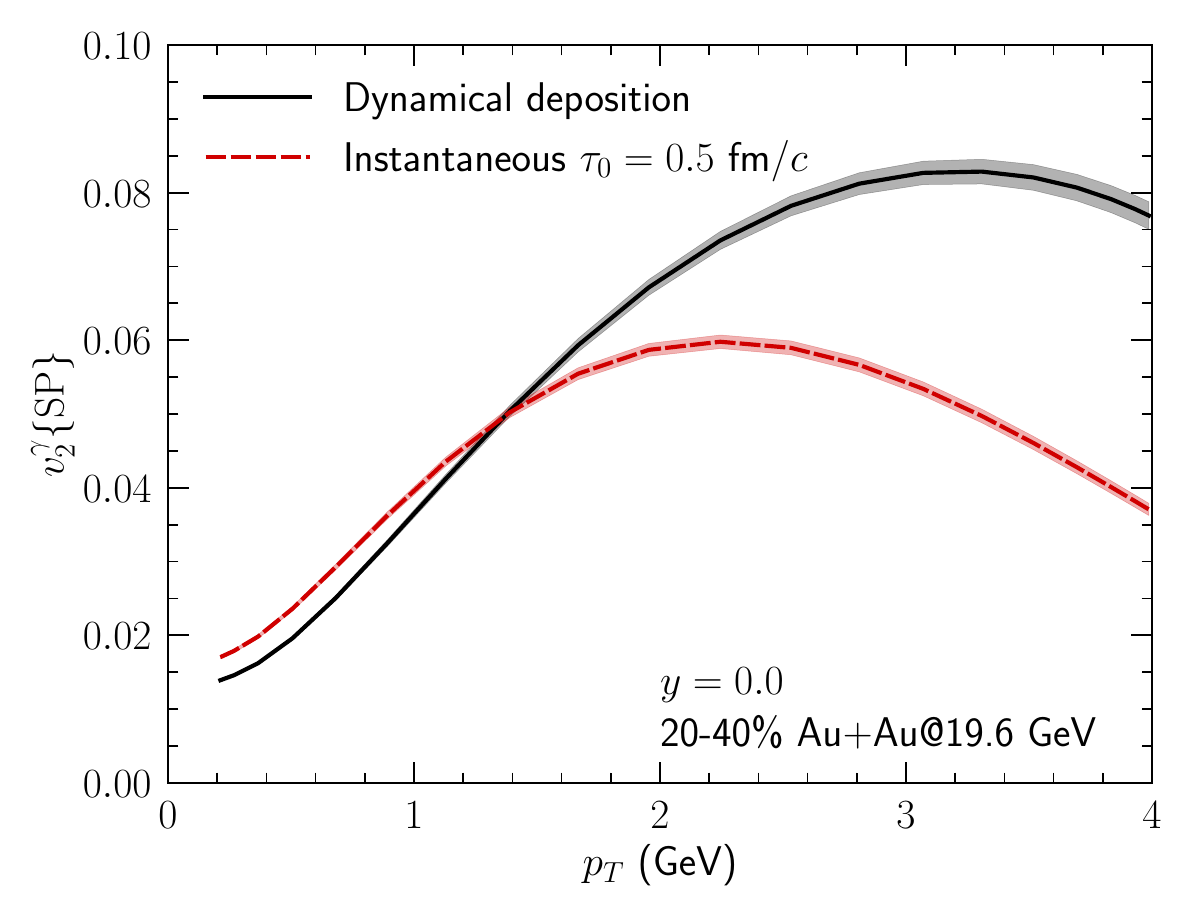}
    \end{tabular}
    \caption{\textit{Top Panels:} Time evolution of averaged fireball temperature and momentum anisotropy in Au+Au collisions at 19.6 GeV for dynamical deposition and instantaneous hydrodynamization scenarios. \textit{Middle Panels:} Charged hadron $p_T$-spectra and $p_T$-differential $v_2$ in central and semi-peripheral Au+Au collisions. Simulations are compared with the STAR measurements. \textit{Bottom Panels:} The mid-rapidity thermal photon spectra and elliptic flow coefficients in Au+Au collisions at 19.6 GeV.}
    \label{fig:EarlyStageEvo}
\end{figure}

The top panels of Fig.~\ref{fig:EarlyStageEvo} compare the dynamical evolution of the fireball's averaged temperature and momentum anisotropy from the dynamical initialization scheme with those from instantaneous initialization of the hydrodynamic fields at a fixed proper time $\tau_0 \equiv \sqrt{t^2 - z^2} = 0.5$ fm/$c$. With the dynamical initialization, the energy-momentum currents are gradually deposited to hydrodynamic fields in the first 2 fm/$c$. This type of dynamics results in the averaged temperature increasing from 0 and reaching a peak value of around 240 MeV in central Au+Au collisions at 19.6 GeV. Compared to the instantaneous initialization, the system's maximum temperature is significantly lower as the $T_\mathrm{max}$ in the latter case is artificially determined by the hydrodynamics starting time $\tau_0$. In the instantaneous initialization case, because the entire fireball is present at $\tau_0 = 0.5$ fm/$c$, it can quickly generate a fast expansion, leading to a faster cooling than the dynamical initialization case. % at the late time.
The dynamical initialization also delays the development of the system's momentum anisotropy. As the energy-momentum source terms gradually feed into hydrodynamic fields, the  fireball's global spatial eccentricity only forms after 2 fm/$c$ when most source terms are deposited into the fluid. It results in the system's momentum anisotropy only starting to develop after $\tau = 2$\,fm/$c$. Because the system expands slower in the dynamical initialization, the system's momentum anisotropy at the late time is smaller than the one from instantaneous initialization. 

\subsection{Photon production at finite baryon density}

The mid-rapidity fireball created in Au+Au collisions at $\sqrt{s_\mathrm{NN}} = 19.6$\,GeV can reach a  sizable net baryon chemical potential, $\langle \mu_B \rangle \sim 200$\,MeV. Therefore, the thermal photon emission will be enhanced by the fluid cell's local net baryon chemical potential $\mu_B$. In the QGP phase, we compute the $\mu_B$ enhancements for the $2 \rightarrow 2$ scattering processes between quarks and gluons~\cite{Gale:2018vuh}. The $\mu_B$ dependence in the effective $1 \rightarrow 2$ processes were computed and parameterized in Ref.~\cite{Gervais:2012wd}. In the hadronic phase, we use the parameterization in Ref.~\cite{Heffernan:2014mla} with the $\mu_B$ dependence. We switch from the QGP rates to hadronic photon production rates at a fixed temperature $T_\mathrm{sw} = 155$ MeV. A systematic exploration of the $T_\mathrm{sw}$ will be reported in future work. For central Au+Au collisions at 19.6 GeV, we found a 10-20\% enhancement of thermal photon production from the finite $\mu_B$ enhancements in the photon production rates. 

\section{Probe the Dynamics of Heavy-ion Collision with Hadrons and Photons}
\label{Sec:Results}

Figure~\ref{fig:EarlyStageEvo} shows the effects of dynamical initialization on hadron (middle panels) and photon observables (bottom panels) by comparing its results with those from the instantaneous initialization. Although there are sizable differences in the system's averaged temperature and momentum anisotropy evolution, the charged hadron $p_T$-spectra and elliptic flow coefficients show small sensitivity to the delay of hydrodynamic flow development in the dynamical initialization scenario. The main reason is that most of the hadron momenta are frozen out late in evolution. They are only sensitive to the distribution of $T^{\mu\nu}$ at the late stage when the hydrodynamic evolution has erased a large extent of the dynamical differences generated at the early time between initialization setups.

On the contrary, thermal photons are more sensitive to the fireball's early-time evolution than charged hadrons because they are continuously produced through the collision evolution with a $T^4$ enhanced weight toward the early stage. Because electromagnetic interactions are much weaker than strong interactions, the photons' momenta are preserved after their production, offering a clean probe of the early-time dynamics of the system.
We find that the high-$p_T$ photon emission is suppressed for dynamical initialization because of the lower maximum temperature compared to the instantaneous initialization. The reduced number of high $p_T$ photons from the early stage also leads to a larger elliptic flow $v_2(p_T)$ for the dynamical initialization. This is because most early-time photons carry small momentum anisotropy, as the hydrodynamic flow anisotropy has not yet developed. 
The thermal photons with transverse momenta below 1.5 GeV mostly come from the late-stage hadronic phase. For these  photons, the dynamical initialization results in smaller elliptic flow compared to those from instantaneous initialization, in line with the system's development of momentum anisotropy at late time. 

The comparison between the middle and bottom panels of Fig.~\ref{fig:EarlyStageEvo} shows that thermal photon observables are much more sensitive to the early-stage dynamics of heavy-ion collisions than hadronic observables. 
We find that such strong sensitivity in photon observables is not significantly diluted after adding the contribution from prompt photon production. As the collision energy goes down, our estimates of prompt photon production suggest that they decrease faster than the thermal radiation at high $p_T$ ($p_T \gtrsim 2-3$~GeV). Therefore, the thermal photon signal can still shine over the prompt background in low-energy heavy-ion collisions.

\section{Conclusions}
\label{Sec:Conclusion}

In this work, we developed a comprehensive theoretical framework to study the event-by-event photon and hadron production in (3+1)D relativistic heavy-ion collisions in the RHIC BES energy region. We show that photon observables are sensitive to the early-time dynamics of Au+Au collisions, providing complementary information to the hadronic observables. By performing a combined phenomenological analysis with hadron and photon measurements, we can derive strong constraints on the dynamical evolution of heavy-ion collisions in a baryon-rich environment and extract how the QGP transport properties depend on the net baryon chemical potential $\mu_B$.

\bigskip

\noindent \textbf{Acknowledgments} 
This work was supported in part by the Natural Sciences and Engineering Research Council of Canada, and in part by the National Science Foundation under Grant No. NSF PHY-1748958, and in part by the US Department of Energy under Contract No.~DE-SC0012704, and Award No.~DE-SC0021969. The numerical simulations used computing resources provided by the Open Science Grid (OSG), supported by the National Science Foundation award \#2030508.
C.S. acknowledges support from a DOE Office of Science Early Career Award.

\bibliographystyle{JHEP}
\bibliography{references}

%\begin{thebibliography}{99}
%\bibitem{...}
%....

%\end{thebibliography}

\end{document}